\documentstyle[12pt]{article}
minus.5em \abovedisplayshortskip=2mm plus.2em minus.4em
\def\title#1#2#3#4#5{
\begin{center} \begin{tabular}[t]{l} #1 \end{tabular} \hfill
\begin{tabular}[t]{r} #2 \end{tabular} \\[1cm] {\LARGE\bf #3} \\[.5in] {#4{}}
\end{center} \vfill \centerline{{\large ABSTRACT}} {\nopagebreak
\noindent\begin{quotation}\noindent {\small #5}

\end{quotation}} \vfill \newpage
\def\thefootnote{\sharp\arabic{footnote}}}

\begin{document}

\thispagestyle{empty}
\title {July 1995} {\bf OSU Preprint 304} {\bf A new ansatz:
Fritzsch Mass Matrices with least modification
} {\ {\bf B. Dutta and S. Nandi }\\
Department of Physics\\Oklahoma State University\\ Stillwater, OK 74078\\}
 {We investigate how
the Fritzsch ansatz for the quark mass matrices can be modified in the least
possible way to accommodate the observed large top quark mass and the
measured values of the CKM elements. As one of the solutions, we find that
the \{23\} and the \{32\} elements of the up quark mass matrix are unequal.
The rest of the assumptions are same as in Fritzsch ansatz. In this
formalism we have an extra parameter i.e. the ratio of the \{{23\}} and the
\{{32\}} element, which gets fixed by the large top quark mass. The
predicted values for $\frac{V_{ub}}{V_{cb}}$ , $\frac{V_{td}}{V_{ts}}$ from
this new ansatz are in the correct experimental range even for the smaller
values of $\tan \beta $. In the end, we write down the $SO(10)$ motivated
superpotential for these new mass matrices } A simple approach for including
the observed hierarchy of quark masses and mixing angles was suggested by
Fritzsch\cite{[fr]}. It prescribes a form for the mass matrices which has
certain amount of predictive power. It can predict the top mass based on the
$V_{cb}$ element of the CKM matrix as input. The top mass according to this
ansatz can not be more than 90 GeV. The recent CDF data\cite{[cdf]} however
shows that the top mass (pole mass) is above 160 GeV ($m_t=176\pm 8\left(
stat\right)\pm 10\left(syst\right) $GeV). So, the simplest
version is clearly excluded.

This situation improves if one realizes these mass matrices at the GUT scale%
\cite{[babu]}. The effect of the running from the weak scale upto GUT scale
is utilized to incorporate a somewhat heavier top in the theory. But the
improvement is not enough to incorporate the recently discovered heavy top
quark. If we try to use this top mass as input in this GUT scenario, the
predicted value of $V_{cb}$ is far off from the experimentally predicted
range which is $V_{cb}=0.0400\pm 0.0025\pm 0.0020$\cite{[vcb]}. In Fig.1 we
show the range of $V_{cb}$ values for the range of the top mass $m_t$ from
155 to 185 GeV , where $m_t$ stands for the running mass. The relation
between the running mass and the pole mass is given by
\begin{equation}
\label{one}m_t=m_t\left( m_t\right) \left[ 1+\frac 43\frac{\alpha _3}\pi
\right]
\end{equation}

Increasing the value of $\tan \beta $ improves the situation a little better
as shown in Fig.1. However, to get the correct prediction for $V_{cb}$, one
has to go beyond $\tan \beta \approx 65$, where the theory loses the
perturbative nature \cite{[barger]}.

In this letter we propose a modification of this Fritzsch ansatz in the
least possible sense. We make the \{23\} and the \{32\} element asymmetrical
in the up quark mass matrix. It then looks like:
\begin{equation}
\label{MU}M_u=P_uUQ_u
\end{equation}
where%
$$
U=\left(
\begin{array}{ccc}
0 & a_u & 0 \\
a_u & 0 & b_u \\
0 & b_u^{^{\prime }} & c_u
\end{array}
\right)
$$

The zeros are described later in the superpotential. The expressions $%
a,b,b^{^{\prime }},c$ are real numbers while $P$ and $Q$ are diagonal phase
matrices. Among the quark phases contained in $P$ and $Q$, only two are
relevant for quark mixing, we denote them by $\psi $ and $\phi $ .The down
matrix is kept the same as in ref.\cite{[fr]}, i.e.
\begin{equation}
\label{MD}M_d=P_dDQ_d
\end{equation}
where%
$$
D=\left(
\begin{array}{ccc}
0 & a_d & 0 \\
a_d & 0 & b_d \\
0 & b_d & c_d
\end{array}
\right)
$$
We also observe the same hierarchical form as has been exercised in the
Fritzsch ansatz, i.e. $c\gg b\sim b^{^{\prime }}\gg a.$ We realize these
matrices at the GUT scale.

The real matrix $U$ is diagonalized by the bi-orthogonal transformation $%
R_uUR_u^{^{\prime }-1}=U^{diag}$ producing the eigenvalues \{$m_u,-m_c,m_t$%
\}.Using the hierarchy of masses $m_u\ll m_c\ll m_t$ we obtain%
$$
R_u=\left(
\begin{array}{ccc}
1 & s_1^u-\chi _1^us_2^u & s_1^us_2^u+\chi _1^u \\
-s_1^u & 1 & s_2^u \\
-\chi _1^u & -s_2^u & 1
\end{array}
\right)
$$

$$
R_u^{^{\prime }}=\left(
\begin{array}{ccc}
1 & s_1^u-\chi _2^us_2^{^{\prime }u} & s_1s_2^{^{\prime }u}+\chi _2^u \\
-s_1^u & 1 & s_2^{^{\prime }u} \\
-\chi _2^u & -s_2^{^{\prime }u} & 1
\end{array}
\right)
$$

where $s_1^u\equiv \sin \varphi _1^u=\sqrt{\frac{m_u}{m_c}}$ and s$%
_2^u\equiv \sin \varphi _2^u=-\sqrt{\frac{m_c}{m_t}}$ and we have set $\cos
\varphi _i^u\approx 1.$ Similarly $\chi _1^u\equiv \frac{m_cs_1^us_2^u}{rm_t}%
,$ $\chi _2^u=\chi _1^ur$ , $s_2^{^{\prime }u}=\frac{s_2^u}r,$ and $r\equiv
\frac b{b^{^{\prime }}}$.

The down quark mass matrix $D$ is diagonalized by the orthogonal
transformation $R_dDR_d^{-1}=D^{diag}$ producing the eigenvalues \{$%
m_d,-m_s,m_b$\}. Using the heirarchy of masses $m_d\ll m_s\ll m_b$ , we
obtain%
$$
R_d=\left(
\begin{array}{ccc}
1 & s_1^d-\chi _1^ds_2^d & s_1^ds_2^d+\chi _1^d \\
-s_1^d & 1 & s_2^d \\
-\chi _1^d & -s_2^d & 1
\end{array}
\right)
$$
$s_1^d\equiv \sin \varphi _1^d=\sqrt{\frac{m_d}{m_s}}$ and $s_2^d\equiv \sin
\varphi _2^d=-\sqrt{\frac{m_s}{m_b}}$ and $\chi _1^d=\frac{m_ss_1^ds_2^d}{m_b%
}$. V$_{CKM}$ at the unification scale in terms of the mass ratios is given
by:
\begin{equation}
\label{vck}R_u\left(
\begin{array}{ccc}
1 &  &  \\
& e^{i\sigma } &  \\
&  & e^{i\tau }
\end{array}
\right) R_d^{-1}
\end{equation}
We find the expressions for $V_{cb}^0,V_{ub}^0,V_{td}^0$ $,V_{ts}^0$ as :
\begin{equation}
\label{vcb}\left| V_{cb}^0\right| =\left| \sqrt{\frac{m_s}{m_b}}-e^{i\phi }%
\sqrt{\frac{m_cr}{m_t}}\right|
\end{equation}
where $\phi =\tau -\sigma ,$%
\begin{equation}
\label{vub}\left| V_{ub}^0\right| =\left| \frac{m_s}{m_b}\sqrt{\frac{m_d}{m_b%
}}+e^{i\psi }\left( \sqrt{\frac{m_s}{m_b}}\sqrt{\frac{m_u}{m_c}}-e^{i\phi
}\left( \sqrt{\frac{m_ur}{m_t}}-\frac{m_c}{m_t}\sqrt{\frac{m_u}{m_tr}}%
\right) \right) \right|
\end{equation}
where $\psi =\sigma ,$%
\begin{equation}
\label{vtd}\left| V_{td}^0\right| =\left| -\frac{m_c}{m_t}\sqrt{\frac{m_u}{%
m_tr}}-e^{i\psi }\sqrt{\frac{m_d}{m_s}}\left( \sqrt{\frac{m_s}{m_b}}%
-e^{i\phi }\sqrt{\frac{m_cr}{m_t}}+\left( \frac{m_s}{m_b}\right) ^{\frac
32}\right) \right|
\end{equation}
\begin{equation}
\label{vts}\left| V_{ts}^0\right| =\left| \sqrt{\frac{m_cr}{m_t}}-e^{i\phi }%
\sqrt{\frac{m_s}{m_b}}\right|
\end{equation}
The zeros in the superscript indicates the GUT value. All the masses are at
the GUT scale ($\sim 10^{16}$). We run them to the top scale ($\sim 170GeV$%
), and also assume that the M$_{SUSY}=m_t$.

To accomplish the running, we write down the Yukawa sector which has the
generic form

\begin{equation}
{\cal L}_Y = \bar{q}_L H_u \phi_u u_R + \bar{q}_L H_d \phi_d d_R + \bar{\ell}%
_L H_\ell \phi_d e_R + h.c.
\end{equation}

\noindent
where $H_u,H_d,H_\ell $ denote the $3\times 3$ Yukawa coupling matrices for
the up quarks, down quarks and charged leptons. The one loop evolution
equations for the Yukawa matrices take the form $(t\equiv \ell n(\mu /M_G)$)%
\cite{[mv]}:
\begin{eqnarray}
16 \pi^2 {{dH_u}\over {dt}} & = & \left[ Tr (3 H_u H_u^{\dagger} +
3a H_d H_d^{\dagger}
+ a H_\ell H_\ell^{\dagger})\right.\nonumber \\
& & + \left. \frac{3}{2} (b H_u H_u^{\dagger} + c H_d H_d^{\dagger}) -
G_U \right] H_u\strut \nonumber \\
16 \pi^2 {{dH_d}\over {dt}} & = & \left[ Tr (3a H_u H_u^{\dagger} +
3H_d H_d^{\dagger} + H_\ell H_\ell^{\dagger}) \right. \nonumber \\
& & + \left. \frac{3}{2} (b H_d H_d^{\dagger} + c H_u H_u^{\dagger}) -
G_D \right]H_d\strut\\
16\pi^2 {{dH_\ell}\over {dt}} & = & \left[ Tr (3a H_u H_u^{\dagger} + 3 H_d
H_d^{\dagger} + H_\ell H_\ell^{\dagger}) \right.\nonumber \\
& & + \left. \frac{3}{2} b H_\ell H_\ell^{\dagger} -
G_E \right] H_\ell \nonumber ~~.
\end{eqnarray}
For the minimal supersymmetric standard model (MSSM) under consideration the
coefficients $a,b,c$ are given by

\begin{equation}
\begin{array}{lcll}
(a,b,c) & = & (0,2,\frac 23) & \rule[-.5cm]{0cm}{1cm}
\end{array}
\end{equation}

\noindent
and the quantities $G_U,G_D$ and $G_E$ are given by:

\begin{equation}
\begin{array}{lcl}
G_U & = & \frac{13}{15}g_1^2+3g_2^2+\frac{16}3g_3^2\rule[-.5cm]{0cm}{1cm} \\
G_D & = & \frac 7{15}g_1^2+3g_2^2+
\frac{16}3g_3^2\rule[-.5cm]{0cm}{1cm} \\ G_E & = & \frac 95g_1^2+3g_2^2~~;
\end{array}
{}~~~~
\end{equation}
\vspace{.2in}

The gauge couplings $g_i$ (above) obey the standard one loop renormalization
group equations:

\begin{equation}
8 \pi^2 \frac{dg^2_i}{dt} = b_i g^4_i,~~~~ i = 1,2,3
\end{equation}

\noindent
where

\begin{equation}
\begin{array}{lcll}
(b_1,b_2,b_3) & = & (\frac{33}5,1,-3) & for {\rm MSSM}%
\rule[-.5cm]{0cm}{1cm}
\end{array}
\end{equation}
\noindent
{}From eqn.(9), one can compute the evolution equations for the eigenvalues
of the Yukawa coupling matrices\cite{[ma]}\cite{[ba]}:

\begin{eqnarray}
16\pi^2 {{df_i}\over {dt}} & = & f_i \left[ 3 \sum_{j=u,c,t} f^2_j + 3a
\sum_{\beta=d,s,b} f^2_\beta + a \sum_{b=e,\mu,\tau} f^2_b - G_U \right.
\nonumber \\
& & \left. + \frac{3}{2} bg^2_i + \frac{3}{2} c \sum_{\beta=d,s,b}
f^2_\beta \mid
V_{i\beta} \mid^2 \right]\strut \nonumber \\
16\pi^2 {{df_\alpha}\over {dt}} & = & f_\alpha \left[ 3a \sum_{j=u,c,t}
f^2_j + 3 \sum_{\beta=d,s,b} f^2_\beta + \sum_{b=e,\mu,\tau} f^2_b -
G_D \right. \nonumber \\
& & \left. +  \frac{3}{2} b f^2_\alpha + \frac{3}{2} c \sum_{j=u,c,t} f^2_j
\mid
V_{j\alpha} \mid ^2 \right]\strut \nonumber \\
16\pi^2 {{df_a}\over {dt}} & = & f_a \left[ 3a \sum_{j=u,c,t} f^2_j + 3 \right.
\sum_{\beta=d,s,b} f^2_\beta + \sum_{b=e,\mu,\tau} f^2_b - G_E \nonumber
\\
& & + \left. \frac{3}{2} b f^2_a \right]
\end{eqnarray}
where $i=(u,c,t),~\alpha =(d,s,b),~a=(e,\mu ,\tau )$.

We will also need the evolution equations for the elements of the CKM matrix%
\cite{[ma]}\cite{[ba]}:
\begin{eqnarray}
16\pi^2 {{d}\over {dt}} \mid V_{i\alpha} \mid^2 & = & 3c \left[
\sum_{j \neq i}\sum_{\beta=d,s,b} {{f^2_i + f^2_j}\over{f^2_i-f^2_j}
} f^2_\beta
Re \left( V_{i\beta} V^*_{j\beta} V_{j\alpha}
V^*_{i\alpha} \right)\strut \right. \nonumber \\
& & \left. + \sum_{\beta\neq \alpha}\sum_{j=u,c,t} {{f^2_\alpha +
f^2_\beta}\over {f^2_\alpha - f^2_\beta}} f^2_j Re \left( V^*_{j\beta}
V_{j\alpha} V_{i\beta} V^*_{i\alpha} \right)\right]~~.
\end{eqnarray}

The above expressions would simplify considerably if we exploit the
hierarchy in the Yukawa couplings ($f_b\gg f_s\gg f_d$, etc) and in the CKM
matrix elements. If only the leading terms are kept, one obtains the
following approximate expressions for the evolution of the various mass
ratios and the mixing angles:
\begin{eqnarray}
16\pi^2 {{d}\over {dt}} \left( {{m_\alpha}\over {m_b}} \right) & = &
-\frac{3}{2} \left( {{m_\alpha}\over {m_b}} \right) \left( b f_b^2 + c
f^2_t), ~~~~\alpha = d,s \right.\strut \nonumber \\
16\pi^2 {{d}\over {dt}} \left( {{m_i}\over {m_t}} \right) & = &
-\frac{3}{2} \left( {{m_i}\over {m_t}} \right) (b f^2_t + c f^2_b),
{}~~~~i =
u,c\strut \nonumber \\
16\pi^2 {{d}\over {dt}} \left({{m_d}\over {m_s}}\right) & = & -\frac{3}{2}
\left({{m_d}\over {m_s}}\right) \left( b f^2_s + c f^2_c + c f^2_t \mid
V_{ts} \mid^2 \right)\strut \nonumber \\
16\pi^2 {{d}\over {dt}} \left( {{m_u}\over {m_c}} \right) & = & -
\frac{3}{2} \left({{m_u}\over {m_c}} \right) \left( b f^2_c + c f^2_s +
c f^2_b \mid V_{cb} \mid^2 \right)\strut \nonumber \\
16\pi^2 {{d}\over {dt}} \mid V_{i\alpha} \mid & = & - \frac{3}{2} c \mid
V_{i\alpha} \mid \left( f^2_t + f^2_b \right)\;~~~~ (i\alpha) = (ub), (cb),
(td), (ts)\strut \nonumber \\
16\pi^2 {{d}\over {dt}} \mid V_{us} \mid & = & - \frac{3}{2} c \mid
V_{us} \mid \left( f^2_c + f^2_s + f^2_t {{\mid V_{td}\mid^2 - \mid
V_{ub} \mid^2}\over {\mid V_{us} \mid^2}} \right)\strut \nonumber \\
16\pi^2 {{d}\over {dt}} \mid V_{cd} \mid & = & - \frac{3}{2} c \mid
V_{cd} \mid \left( f^2_c + f^2_s + f^2_b {{\mid V_{ub}\mid^2 -
\mid V_{td} \mid^2}\over {\mid V_{cd}\mid^2}}\right)~~.
\end{eqnarray}

\noindent
One comment is necessary here:

$\frac{\left| V_{ub}\right| }{\left| V_{cb}\right| }$ and $\frac{\left|
V_{td}\right| }{\left| V_{ts}\right| }$ do not run in one loop.

\noindent
The low energy i.e. the m$_t$ scale values of the Yukawa couplings are :
\begin{equation}
\label{lambda}\lambda _b(m_t)=\frac{\surd 2m_b(m_b)}{\eta _bv\cos \beta }%
,\lambda _\tau (m_t)=\frac{\surd 2m_\tau (m_\tau )}{\eta _\tau v\cos \beta }%
,\lambda _t(m_t)=\frac{\surd 2m_t(m_t)}{v\sin \beta }
\end{equation}
where $\eta _f$ =$m_f(m_f)/m_f(m_t)$ gives the running of the masses below $%
\mu =m_t,$ obtained from 3-loop QCD and 1 loop QED evolution, for heavy
flavors $f=t,b,c,\tau .$ For light flavors $f=s,d,e,\mu $ we stop at $\mu =$%
1 GeV and define $\eta _f$ =$m_f(1GeV)/m_f(m_t)$ . For $\alpha _3=0.118,$ $%
\eta _b\simeq 1.5,$ $\eta _c\simeq 2.1,\eta _s=\eta _d=\eta _u\simeq 2.4.$
The running mass values are $m_b(m_b)=4.25\pm 0.15GeV,m_\tau (m_\tau
)=1.7777GeV,m_c(m_c)=1.2GeV,m_s(1GeV)\simeq 0.175GeV,m_u(1GeV)\simeq
0.006GeV,m_d(1GeV)\simeq 0.008GeV$\cite{[gasser]}.

We solve for $r\left( =\frac b{b^{^{\prime }}}\right) $ from $V_{cb}^0$
using eqn.(\ref{vcb}) for a range of values of $\varphi $ as shown in Table 1.
The values of $\varphi $ are chosen so that $r$ is real. We use these values of
$r$ to predict $V_{ub}^0$(eqn.(%
\ref{vub})),$V_{td}^0$(eqn.(\ref{vtd})) and $V_{ts}^0$(eqn.(\ref{vts})). We
then
calculate the values of $V_{ub}$, $V_{cb}$, $V_{td}$ and  $V_{ts}$ at the low
scale using eqn.(17). In
Table \ref{one} , we present the values of $\frac{V_{ub}}{V_{cb}}$ and $\frac{%
V_{td}}{V_{ts}}$ from our ansatz and compare them with the experimental
values. The dependence of any prediction on $\psi $ is negligible, $\psi $
is kept fixed at $\pi /2$. As shown in the Table 1, the prediction of the model
is in excellent agreement with the current experimental ranges for an wide
range
of $\tan\beta$. The model will be tested further as the experimental ranges
are narrowed down in the future.
\\\indent Motivated by the supersymmetric
$SO(10)$ Grand Unifying group, we can write down the
superpotential for the mass matrices
\begin{equation}
\label{sup}
\begin{array}{lll}
W & = & f_{12}^{(10^{^{\prime \prime }})}\psi _1\psi _2\varphi
_{10^{^{\prime \prime }}}+f_{23}^{(10^{^{\prime }})}\psi _2\psi _3\varphi
_{10^{^{\prime }}}+f_{23}^{(120)}\psi _2\psi _3\varphi
_{120}+f_{33}^{10}\psi _3\psi _3\varphi _{10}+h.c.
\end{array}
\end{equation}
Here, $\varphi _{120}$ has the vev only in the up direction.
The zeros are produced by the discrete symmetry. From eqn.(19), we
obtain the following form of the mass matrices.

\begin{equation}
\label{ud}U=\left(
\begin{array}{ccc}
0 & 10^{^{\prime \prime }} & 0 \\
10^{^{\prime \prime }} & 0 & 10^{^{\prime }},120 \\
0 & 10^{^{\prime }},120 & 10
\end{array}
\right) ,D=\left(
\begin{array}{ccc}
0 & 10^{^{\prime \prime }} & 0 \\
10^{^{\prime \prime }} & 0 & 10^{^{\prime }} \\
0 & 10^{^{\prime }} & 10
\end{array}
\right)
\end{equation}
Here, the entries in an element correspond to the Higgs
fields contributing to that element.
Since \{120\} is an antisymmetric representation , the \{23\} and \{32\}
elements are asymmetric.

In conclusion we summarize that if \{23\} and \{32\} elements of the
up quark sector of the Fritzsch mass matrices are unequal, one can predict
$\frac{V_{ub}}{V_{cb}}$ and $\frac{%
V_{td}}{V_{ts}}$ in the correct experimental range even with a heavy top in
the theory. Moreover, for this ansatz, $\tan \beta $ need not be very high,
it can lie anywhere between 0.6 and 65. Also, it is possible to write a
superpotential for this ansatz.\\
This research was supported in part by the US Department of Energy,
Grant Number DE-FG02-94ER40852.

{\bf TABLE CAPTION}

\begin{quote}
{\bf Table 1}\\ The predicted values of $\frac {V_{ub}}{V_{cb}} $ $\frac
{V_{td}}{V_{ts}} $ from the new ansatz are compared with their experimental
values for different values of $\tan \beta$ and $\tan \phi $
\end{quote}

{\bf FIGURE CAPTION}

\begin{quote}
{\bf Figure 1}\\The predicted values of $V_{cb}$ from Fritzsch ansatz is
plotted as functions of $m_t $, for values of $\tan\beta$=3 and 60.
The experimental range of $V_{cb}$ is also
shown. To be conservative, we have taken the deviation from the central value
of $V_{cb}$ to be 0.006 instead of 0.003.\newpage\
\begin{table} \centering
\caption{\label{key}}
  \begin{tabular}{|l|l|l|l|l|l|} \hline
     {\bf tan $\beta $}  &  {\bf tan $\phi $}  &  {\bf predicted $\frac
{V_{ub}}{V_{cb}}$} & {\bf predicted $ \frac { V_{td}}{V_{ts}}$}
&{\bf Exptal.range of $\frac {V_{ub}}{V_{cb}}$} &{\bf Exptal.range of $\frac
{V_{td}}{
V_{ts}}$} \\ \hline
            &    0    & 0.069  & 0.23 & & \\
        3     &   0.18 & 0.082 & 0.23 & 0.03-0.137 &0.11-0.36 \\
            &    0.23     &  0.084 & 0.22 &  & \\\cline{1-4}
           &  0       &  0.069 & 0.23 & & \\
                  20 & 0.18  & 0.080   &  0.22   & & \\
                     & 0.24 &  0.084  &  0.22         &  & \\\cline{1-4}
                  &     0  & 0.068  & 0.23 & &  \\
            40  &    0.18   & 0.078 & 0.23 & & \\
                  &    0.24 & 0.082 & 0.22 & & \\\cline{1-4}
             &  0  & 0.067 & 0.22 & & \\
              60    &  0.18 & 0.072 & 0.22 & &\\
                  &  0.26 & 0.075 & 0.22 & & \\\hline
  \end{tabular}

\end{table}
\end{quote}

\end{document}